\documentclass[aoas]{imsart}

\RequirePackage{amsthm,amsmath,amsfonts,amssymb}
\RequirePackage[authoryear]{natbib}
\RequirePackage[colorlinks,citecolor=blue,urlcolor=blue]{hyperref}
\RequirePackage{graphicx}
\usepackage{algorithm}
\usepackage{algorithmic}
\usepackage[bblfile=\jobname]{biblatex-readbbl}

\startlocaldefs
\theoremstyle{plain}

\theoremstyle{remark}

\endlocaldefs

\begin{document}

\begin{frontmatter}
\title{Multi-View Orthogonal Projection Regression with Application in Multi-Omics integration}
\runtitle{A sample running head title}

\begin{aug}
\author[A]{\fnms{Zongrui}~\snm{Dai} \ead[label=e1]{daizr@umich.edu}\orcid{0000-0002-7893-5004}},
\author[B]{\fnms{Yvonne}~\snm{J. Huang}\ead[label=e3]{yvjhuang@med.umich.edu}\orcid{0000-0002-7497-6597}}
\and
\author[A]{\fnms{Gen}~\snm{Li}\thanks{[\textbf{Corresponding author indication should be put in the Acknowledgment section if necessary.}]}\ead[label=e2]{ligen@umich.edu }\orcid{0000-0002-7298-2141}}
\address[A]{Department of Biostatistics, University of Michigan, Ann Arbor,Michigan\printead[presep={ ,\ }]{e1,e2}}

\address[B]{Department of Internal Medicine, Division of Pulmonary and Critical Care Medicine, Department of Microbiology and Immunology, University of Michigan
Medical School, Ann Arbor, Michigan\printead[presep={,\ }]{e3}}
\end{aug}

\begin{abstract}
Multi-omics integration offers novel insights into complex biological mechanisms by utlizing the fused information from various omics datasets. However, the inherent within- and inter-modality correlations in multi-omics data present significant challenges for traditional variable selection methods, such as Lasso regression. These correlations can lead to multicollinearity, compromising the stability and interpretability of selected variables. To address these problems, we introduce the Multi-View Orthogonal Projection Regression (MVOPR), a novel approach for variable selection in multi-omics analysis. MVOPR leverages the unidirectional associations among omics layers, inspired by the Central Dogma of Molecular Biology, to transform predictors into an uncorrelated feature space. This orthogonal projection framework effectively mitigates the correlations, allowing penalized regression models to operate on independent components. Through simulations under both well-specified and misspecified scenarios, MVOPR demonstrates superior performance in variable selection, outperforming traditional Lasso-based methods and factor-based models. In real-data analysis on the CAARS dataset, MVOPR consistently identifies biologically relevant features, including the \textit{Bacteroidaceae} family and key metabolites which align well with known asthma biomarkers. These findings illustrate MVOPR’s ability to enhance variable selection while offering biologically interpretable insights, offering a robust tool for integrative multi-omics research.
\end{abstract}


\end{frontmatter}

\section{Introduction}
\begin{figure}[H]
\centering\includegraphics[width=12cm]{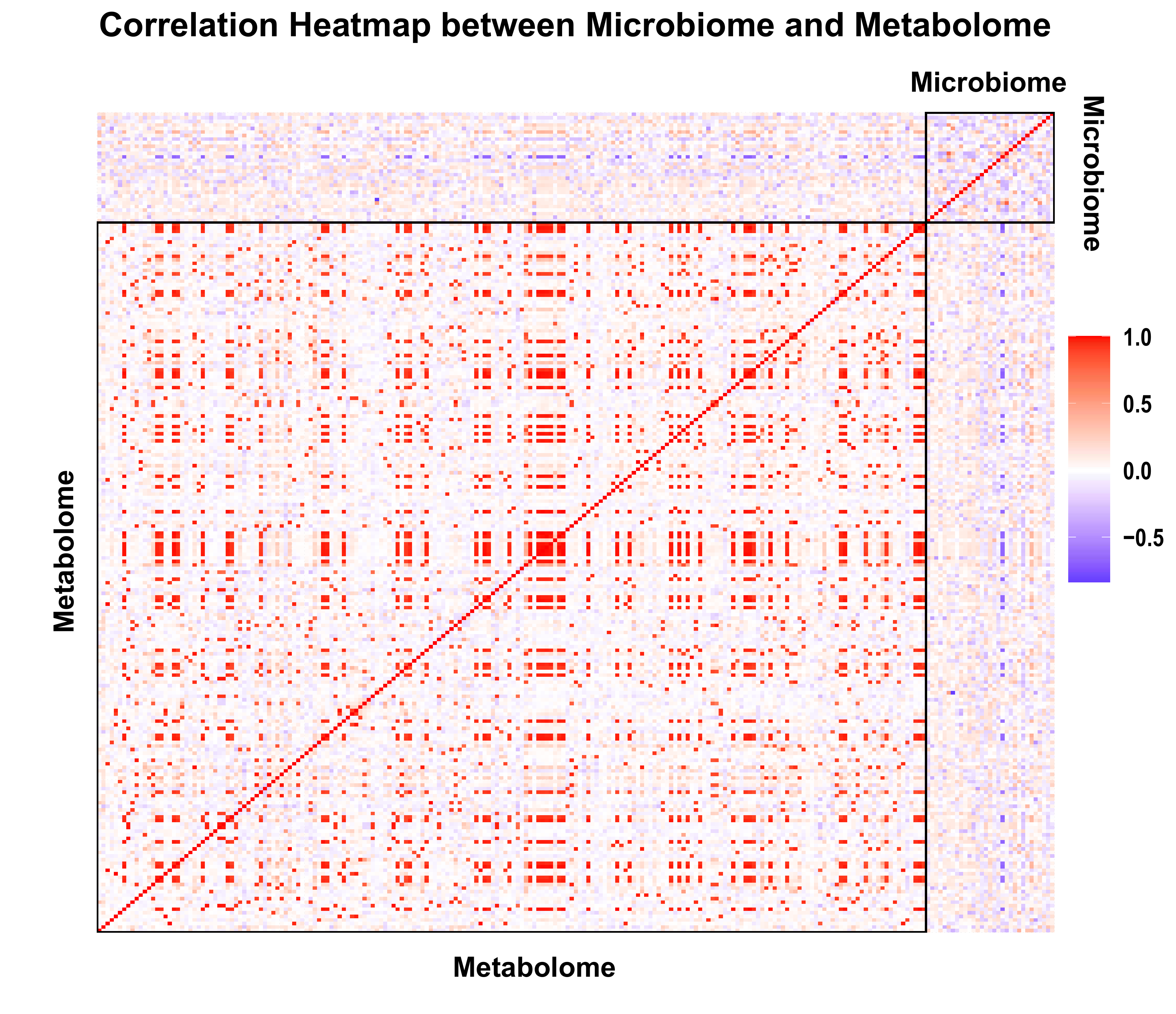}
\caption{The Pearson Correlation Heatmap of Microbiome and Metabolome}
\label{figure.0}
\end{figure}

Multi-omics analysis provides a comprehensive understanding of biological mechanisms by integrating multiple types of data, such as genomics, transcriptomics, proteomics, and metabolomics. These datasets offer a novel insight into molecular processes and immunological research that cannot be found from any single modality alone (\cite{chen2023applications,chu2021multi,clark2021integrative}). Numerous studies have found that the fusion of different omics data can bring additional insights into the exploration of biomarkers, improving diagnostics, and therapy development (\cite{gillenwater2021multi, garg2024disease, olivier2019need, hussein2024multi, menyhart2021multi}). For example, in our recent study of CAARS data, we collected the gut microbiome and metabolome data of 51 patients to investigate the combined impact of these two omics layers on asthma development. \\
Asthma is a complicated respiratory disease which involves airway inflammation and allergic reactions (\cite{gautam2022multi}).  As one of the most prevalent chronic airway diseases, it exhibits high heterogeneity, making diagnosis based on a single biomarker challenging (\cite{chung2016asthma, abdel2020omics}). Increasing evidence suggests that the pathogenesis of asthma is closely linked to different omics data. For instance, potential host-microbiota interactions have been associated with an increased risk of asthma (\cite{ruff2020host} ). Emerging collaborations are using multi-omics data to advance the understanding of asthma. For instance, multi-omics integration has explored the disease's heterogeneity and underlying pathology. This approach not only helps identify new patient stratification, but also paves the way for personalized treatment strategies (\cite{zhang2024unraveling}). \\
Lasso-based regression is a widely used approach for variable selection in multi-omics data. For example, IPF-LASSO (Integrative LASSO with Penalty Factors) and Priority-Lasso are both Lasso-based methods that account for the heterogeneity of multi-omics data by assigning distinct penalty and priority weights in the regression model (\cite{klau2018priority,boulesteix2017ipf}). However, some studies have found that the performance of these models can be influenced when highly correlated predictors are present, as these methods do not account for the within- and inter-modality correlations inherent in multi-omics data (\cite{castel2024comparison}). \\
To illustrate this challenge, we measure the correlation between microbiome and metabolome through the Pearson correlation heatmap in the CAARS dataset. Microbiome data are aggregated to the family level and applied centered log ratio transformation. Metabolome data are centered and scaled. The heatmap shows that metabolome data have strong within-modality correlation compared to the microbiome.  Additionally, there are some negative inter-modality correlations between the microbiome and metabolome (Figure.\ref{figure.0}). 
The within- and inter-modalities correlations pose significant challenges for variable selection in traditional regression models. These correlations arise from latent associations between omics layers and intrinsic dependencies within each individual omic. Network-based approaches provide an effective way to capture these complex relationships, such as protein-protein and protein-RNA interactions (\cite{richards2021mass, nasiri2021novel, szklarczyk2023string}). Although these correlations can be quantified, they introduce complications when modeling the relationship between multi-omics and a response variable. Specifically, these strong correlations can lead to multicollinearity, making it difficult to distinguish the individual contributions of variables. \\
To better justify the problem, suppose there are two modalities $M_{1} \in \mathbb{R}^{n \times p_{1}}$ and $M_{2} \in \mathbb{R}^{n \times p_{2}}$ which are assoicated with response vector $Y$ through $\beta_{1}$ and $\beta_{2}$. A linear model could be constructed below: 
\begin{equation*}
Y = M_{1} \beta_{1} + M_{2} \beta_{2} + \epsilon_{1}
\end{equation*}
The existences of within- and inter-modality correlations indicate the latent relationships of $M_{1}$ and $M_{2}$ below: 
\begin{equation*}
\underbrace{M_{1} = F\Lambda + U}_{\text{Within-Modality Correlation}} 
\quad \text{and} \quad 
\underbrace{M_{2} = M_{1}B + E}_{\text{Inter-Modality Correlation}},
\end{equation*}
where $F$ represents the latent factors in modality $M_{1}$ with rank $r$, and $F\Lambda$ captures the low-rank structure of $M_{1}$, which is a commonly used approach to describe within-modality correlation. For instance, factor-based models frequently employ it to represent dependencies within predictors. The inter-modality correlation shows the $M_{2}$ modality can be represented by the linear combination of $M_{1}$ through the coefficient matrix $B$ with error term $E$. Standard variable selection methods, such as Lasso regression, struggle in this setting because they will arbitrarily select among correlated predictors, potentially leading to unstable variable selection. Addressing this issue requires a method that not only accounts for the relationships between omics layers but also isolates the independent contributions of each omic.\\
The factor model has emerged as a powerful tool for correlated data by decomposing them into latent structures comprising factors and idiosyncratic components. For instance, factor-adjusted regularized regression can handle highly correlated data by identifying and removing the low-rank structure from data and retaining the idiosyncratic components for variable selection(\cite{fan2020factor}). Integrative Factor Regression is another factor decomposition-based model designed for multi-model dataset. It can extract modality-specific factors to account for the heterogeneity across modalities (\cite{li2022integrative}). However, factor-based models require data to have an approximate latent factor structure and only reduce correlations within individual modalities. Thus, correlations between modalities still exist. \\
An alternative approach is cooperative learning, which employs an agreement penalty based on contrastive learning. This method encourages different modalities to contribute similarly. By varying the hyperparameter of the agreement penalty, the solutions for this method include the early and late fusion approaches for multi-model data, providing robust performance for different settings (\cite{ding2022cooperative}). However, this method ignores the inherent correlations between the modalities and enforces different modalities to align the contribution. In multi-omics scenarios, this assumption may not always hold, as different omics layers can have distinct influences on the response variable. For example, miR-155 and miR-146a are well-known miRNAs that can suppress \textit{E. coli}-induced inflammatory responses in neuroinflammation. This example suggests that the host transcriptome may have an opposing effect compared to the microbiome, leading to potential contradictions between the molecular signals originating from the host and those from the microbiota (\cite{yang2021mir}).\\
To address these challenge, we introduce a novel Multi-View Orthogonal Projection Regression (MVOPR) for variable selection in multi-omics data. Unlike existing methods that impose specific structural assumptions on the data, our model leverages unidirectional associations among different omics to mitigate the correlations. Our approach is inspired by the Central Dogma of Molecular Biology, which states that DNA transcribes to RNA, and RNA translates into protein, while the reverse process is impossible. For instance, once a protein is synthesized, it cannot alter its original RNA template. This inherent directionality in molecular interactions suggests that multi-omics relationships can be represented by a directed graph (digraph) with unidirectional pathways. Building on this biological insight, our method accounts for the dependencies by removing redundant correlations in a structured manner. Specifically, we employ an orthogonal projection framework that sequentially remove the effects of upstream omics layers on downstream ones. By transforming the original multi-omics data into an uncorrelated feature space, our approach ensures that variable selection methods, such as penalized regression, operate on independent components free from both within and across modality correlations. This enables MVOPR to overcome the limitations of standard Lasso-based approaches, noted above. \\
In this study, we demonstrate the effectiveness of MVOPR for multi-omics variable selection through both theoretical analysis and empirical validation. Our simulations and real-data analysis reveal that MVOPR consistently outperforms existing methods. We also show that when inter-modality correlation exists, the factor-based models will face different problems, unlike MVOPR. Importantly, even in cases where cross-modality correlations are absent, MVOPR remains robust and performs comparably to standard Lasso regression, demonstrating its adaptability across different correlation structures. By incorporating biological direction assumptions, our approach not only enhances variable selection performance but also aligns with the natural structure of molecular data, offering a robust framework for integrative multi-omics analysis.\\
The rest of the article is organized as follows. In Section 2, we present the MVOPR framework for both the two-modality and multiple-modality scenarios, followed by an introduction to three related methods for multi-modal data analysis. Section 3 provides a comparative analysis of MVOPR against other competing methods under various settings. In Section 4, we apply MVOPR to the CAARS dataset and evaluate its performance relative to alternative approaches.

\section{Methodology}
\subsection{MVOPR for Two Modalities}
Suppose we have two modalities, \( M_{1} \in \mathbb{R}^{n \times p} \) and \( M_{2} \in \mathbb{R}^{n \times q} \). Let \( Y \) be the response vector of length \( n \), assumed to be associated with \( M_{1} \) and \( M_{2} \) through regression coefficients \( \beta_{1} \in \mathbb{R}^{p} \) and \( \beta_{2} \in \mathbb{R}^{q} \). The relationship is modeled as:
\begin{equation}
Y = M_{1} \beta_{1} + M_{2} \beta_{2} + \epsilon_{1},
\end{equation}
where \( \epsilon_{1} \) is an error term assumed to be uncorrelated with \( M_{1} \) and \( M_{2} \), with \( E(\epsilon_{1})=0 \) and \( \operatorname{Var}(\epsilon_{1}) = \sigma^2_{\epsilon_{1}} I \). We further assume that \( M_{2} \) is influenced by \( M_{1} \) through a low-rank coefficient matrix \( B \in \mathbb{R}^{p \times q} \) of rank \( r \), with an error component \( E \in \mathbb{R}^{n \times q} \) that is uncorrelated with \( M_{1} B \) and \( \epsilon_{1} \):
\begin{equation}
M_{2} = M_{1} B + E.
\end{equation}
Using this inter-modality correlation (2), we can reformulate Model (1) as:
\begin{equation}
Y = M_{1} (\beta_{1} + B \beta_{2}) + E\beta_{2} + \epsilon_{1}.
\end{equation}
If \( E \) is small, the model becomes almost unidentifiable, which affects the selection of the variables for \( M_{2} \). To handle this issue, we aim to remove the associated component \( M_{1}B \) from \( M_{2} \) while retaining only the uncorrelated part \( E \).\\
Let \( U \Sigma V^{T} \) be the singular value decomposition (SVD) of \( M_{1}B \), where \( U_{r} \) consists of the first \( r \) left singular vectors of \( U \). Substituting this decomposition into Model (3) gives:
\begin{equation}
Y = M_{1} \beta_{1} + E\beta_{2} + U_{r} \gamma_{1} + \epsilon_{1},
\end{equation}
where \( \gamma_{1} \) is a nuisance parameter. Since \( U_{r} \) captures the principal directions of \( M_{1} B \), it is highly correlated with \( M_{1} \). To eliminate this correlation, we project \( M_{1} \) onto a subspace orthogonal to \( U_{r} \). \\
Define the projection matrix \( P = U_{r} U_{r}^{T} \), and let \( P^{\perp} = I - P \) be its orthogonal complement. Transforming \( M_{1} \) into \( P^{\perp} M_{1} \) ensures that the new predictor no longer lies in the column space of \( M_{1}B \), thereby breaking the correlation with \( U_{r} \). The transformed MVOPR model for two modalities is then:
\begin{equation}
Y = M^{*}_{1} \beta_{1} + M^{*}_{2}\beta_{2} + U_{r} \gamma^{*} + \epsilon_{1},
\end{equation}
where \( M_{1}^{*} = P^{\perp} M_{1} \) and \( M^{*}_{2} = E \). In this formulation, the predictors \( M_{1}^{*} \), \( M_{2}^{*} \), and \( U_{r} \) are mutually uncorrelated.\\
This transformation enhances variable selection for both \( \beta_{1} \) and \( \beta_{2} \) by removing redundant correlations between modalities. The decomposition effectively subtracts the linear contribution of \( M_{1} \) in \( M_{2} \) and projects \( M_{1} \) outside the column space of \( M_{1} B \), ensuring no internal dependencies between \( M_{1}^{*} \), \( M_{2}^{*} \), and the nuisance components \( U_{r} \).
\subsection{MVOPR for Multiple Modalities}
Extending the model to three modalities, let \( M_{1} \in \mathbb{R}^{n \times p_{1}} \), \( M_{2} \in \mathbb{R}^{n \times p_{2}} \), and \( M_{3} \in \mathbb{R}^{n \times p_{3}} \), with a known hierarchical dependency:
\begin{equation*}
M_{2} = M_{1} B_{2,1} + E_{2},
\end{equation*}
\begin{equation*}
M_{3} = M_{1} B_{3,1} + M_{2} B_{3,2} + E_{3}.
\end{equation*}
where \( E_{2} \in \mathbb{R}^{n \times p_{2}} \) and \( E_{3} \in \mathbb{R}^{n \times p_{3}} \) are independent error components, and \( B_{2,1} \), \( B_{3,1} \), and \( B_{3,2} \) are low-rank coefficient matrices with ranks \( r_{1} \), \( r_{2} \), and \( r_{3} \), respectively.\\
The response \( Y \) is modeled as:
\begin{equation}
Y = M_{1} \beta_{1} + M_{2} \beta_{2} + M_{3} \beta_{3} + \epsilon_{1}.
\end{equation}
To remove the associated components, define two projection matrices:  \( P_{1} = U_{1} U_{1}^{T} \) and \( P_{2} = U_{2} U_{2}^{T} \) which based on \( M_{1} B_{1}' \) and \( E_{2} B_{3,2} \) separately. \( B_{1}' = (B_{2,1}, B_{3,1}) \) is the concatenation of \( B_{2,1} \) and \( B_{3,1} \). The transformed modalities are:
\begin{align*}
M_{1}^{*} = (I - P_{1}) M_{1}, \quad M_{2}^{*} = (I - P_{2}) E_{2}, \quad M_{3}^{*} = E_{3}.
\end{align*}
Transforming the modalities in model (6), the MVOPR model for three modalities is:
\begin{equation}
Y  = M_{1}^{*} \beta_{1} + M_{2}^{*} \beta_{2} + M_{3}^{*} \beta_{3} + U_{1} \gamma_{1} + U_{2} \gamma_{2} + \epsilon_{1}.
\end{equation}
where \( \gamma_{1} \) and \( \gamma_{2} \) are nuisance parameters. The detailed derivations are provided in Supplementary Appendix \ref{appA}.\\
For more than three modalities, the transformation follows a similar procedure. Suppose there are \( k \) modalities with features \( p_{1}, p_{2}, \dots, p_{k} \). If each modality \( M_{j} \) (for \( j = 2, 3, \dots, k \)) depends only on previous modalities:
\begin{equation*}
M_{j} = \sum_{i=1}^{j-1} M_{i} B_{j,i} + E_{j},
\end{equation*}
where \( E_{j} \) is independent noise, then the final regression model is:
\begin{equation}
Y = M_{1} \beta_{1} + M_{2} \beta_{2} + \dots + M_{k} \beta_{k} + \epsilon_{1}.
\end{equation}
With the direction assumption above, we could derive the connection between response $Y$ and all the modalities by the following algorithm.

\begin{algorithm}[H]
\caption{Algorithm for multi-view regression on multiple modalities}
\label{alg:sum}

\begin{algorithmic}[1]
    \STATE \textbf{Input:} Multiple modalities $M_{1}, M_{2},...,M_{k}$ and response $Y$
    \STATE \textbf{Step.1:} Obtain the estimation of $B_{2,1},B_{3,1}, ..., B_{k,k-1}$
    \STATE \quad \textbf{for} $j$ in 1:m 
    \STATE \quad \quad Regress $M_{j} \sim  (M_{1},...,M_{j-1})$. Calculate the residuals by $\hat{E}_{j} = M_{j} - \hat{M_{j}}$
    \STATE \textbf{Step.2:} Obtain the projection matrix $P_{1},...,P_{m-1}$
    \STATE \quad Calculate:
    \STATE \quad \quad $M_{1}\hat{B}^{'}_{1} = M_{1}(\hat{B}_{2,1},\hat{B}_{3,1},...,\hat{B}_{k,1})$ 
    \STATE \quad \quad $\hat{E}_{2}\hat{B}^{'}_{2} = \hat{E}_{1}(\hat{B}_{3,2},\hat{B}_{4,2},...,\hat{B}_{k,2})$, ..., $\hat{E}_{k-1}\hat{B}^{'}_{k-1} = \hat{E}_{k-1}\hat{B}_{k,k-1}$. 
    
    \STATE \quad Obtain the SVD: 
    \STATE \quad \quad $M_{1} \hat{B}^{'}_{1} = U_{1} \Sigma_{1} V^{T}_{1}$ and projections $P_{1} = U_{1}U_{1}^{T}$ with rank $r_{1}$. 
    \STATE \quad \quad For $j \geq 2$, $\hat{E}_{j} \hat{B}^{'}_{j} = U_{j} \Sigma_{j} V_{j}^{T}$ and projections $P_{j} = U_{j}U_{j}^{T}$ with rank $r_{j}$. 
    
    \STATE \textbf{Step.3:} Transform the $M_{1}, M_{2},...,M_{m}$ by $M^{*}_{1}, M^{*}_{2},...,M^{*}_{m}$
    \STATE \quad $j=1$: $M^{*}_{1} = P_{1}^{\perp}M_{1}$. 
    \STATE \quad $j = 2,.., k-1$: $M^{*}_{j} = P_{j}^{\perp}\hat{E_{M_j}}$. 
    \STATE \quad $j=k$: $M^{*}_{k} = \hat{E_{M_k}}$
    \STATE \quad Obtain the nuisance variable $U = (U_{1}, U_{2},...,U_{m-1})$
    \STATE \textbf{Step.4:} Obtain the estimation of $\beta_{1},...,\beta_{k}$
    \STATE \quad Solve the penalized optimization:
     \STATE \quad \quad $\min\limits_{\beta,\gamma} \|Y - M_{1}^{*} \beta_{1} -...- M_{k}^{*} \beta_{k} - U \gamma\|^{2} + \sum_{j=1}^{k} P_{\lambda}(\beta_{j})$
    \RETURN $\hat{\beta_{1}},...,\hat{\beta_{k}}$
\end{algorithmic}
\end{algorithm}

\subsection{Related methods}
To evaluate the relative performance of MVOPR, we consider several alternative models for multi-modality data. Specifically, we compare our method against Cooperative Regularized Linear Regression (\textbf{Cooperative}~\cite{Ding2022-og}), Integrative Factor Regression (\textbf{IntegFactor}~\cite{li2022integrative}), and Factor-Adjusted Regularized Regression (\textbf{Factor}~\cite{fan2020factor}).

\subsubsection{Cooperative Regularized Linear Regression}  
Cooperative regularized linear regression is a widely used approach for multi-view learning. It integrates multiple modalities by imposing an agreement penalty that encourages the predictions from different modalities to be aligned. The level of agreement between modalities is controlled by the hyperparameter \( \rho \).  When $\rho=0$, this method becomes traditional penalized regression with chosen penalty. When $\rho=1$, it indicates a late fusion of all the modalities. Suppose there are two modalities $M_{1}$ and $M_{2}$, its least square problem can be written as below: 
\begin{align}
\min_{\beta_{1},\beta_{2}} \| Y - M_{1}\beta_{1} - M_{2}\beta_{2}\|^{2} + \frac{\rho}{2} \| M_{1}\beta_{1} - M_{2}\beta_{2}\|^2 +\lambda_{1} \| \beta_{1} \|_{1} + \lambda_{2} \| \beta_{2} \|_{1} 
\end{align}
To simplify the optimization, $\lambda_{1}$ and $\lambda_{2}$ are equal in this study. Problem (9) is convex, we can transform the original data below: 
\begin{align}
\tilde{X} = \begin{pmatrix} 
M_{1} & M_{2} \\ 
-\sqrt{\rho}M_{1} & \sqrt{\rho}M_{2} 
\end{pmatrix}, \quad 
\tilde{Y} = \begin{pmatrix} 
Y \\ 
0 
\end{pmatrix}, \quad 
\tilde{\beta} = \begin{pmatrix} 
\beta_{1} \\ 
\beta_{2}
\end{pmatrix}.
\end{align}
Based on the transformed data, the problem (9) is equivalent problem to the generic lasso problem below: 
\begin{align}
J(\theta_x, \theta_z) =  \left\| \tilde{Y} - \tilde{X} \tilde{\beta} \right\|^2 + \lambda \left\| \tilde{\beta} \right\|_1 
\end{align}

\subsubsection{Factor-based Models} 
Factor-based methods assume that predictor $M$ follows an approximate factor model 
\begin{equation}
M = F\Lambda + U,
\end{equation}
where $F$ is a $K \times 1$ vector of latent factors, $\Lambda$ is a $p \times K$ loading matrix, and $U$ is the $p \times 1$ vector of idiosyncratic components. By separating the idiosyncratic components from the $M$, it de-correlates the original $M$ to a weakly correlated element $u$. The regression model: 
\begin{equation}
Y = M \beta + \epsilon
\end{equation}
can then be reformulated as:
\begin{equation}
Y = U\beta + F\gamma + \epsilon, \quad \text{where } \gamma = \Lambda\beta \text{ is a nuisance parameter}.
\end{equation}
To estimate the factors \( \hat{F} \) and idiosyncratic components \( \hat{U} \) from \( M \), we adopt the method of Bai and Li (~\cite{bai2012statistical}) and Fan et al.(~\cite{fan2013large}). The optimal number of latent factors \( K \) is selected based on Bai and Ng's information criteria method (~\cite{bai2002determining}). Using these estimates, the least squares problem can be written as follows, where $\gamma$ is the nuisance parameter.
\begin{align*}
\min_{\beta, \gamma} \| Y - \hat{U} \beta - \hat{F}\gamma \|^{2} + \lambda \rho(\beta) 
\end{align*}
In Factor-Adjusted Regularized Regression, the multi-modal data is treated as a unified design matrix, and factor decomposition is applied globally across the entire dataset. Specifically, let $M = (M_{1},...,M_{m})$ represent the concatenation of all modalities (~\cite{fan2020factor}). While Integrative Factor Regression targets multimodal data by modeling each modality separately, allowing for the extraction of modality-specific latent factors and idiosyncratic components. That is, for each modality $M_{i}$ has its own factors $F_{i}$ and idiosyncratic component $U_{i}$ with $i \in 1,...,m$. Then, the regression model is fitted using the concatenated idiosyncratic components and latent factors, where \( U = (U_{1},...,U_{m}) \) and \( F = (F_{1},...,F_{m}) \) represent the concatenation of all modality-specific idiosyncratic components and latent factors, respectively. \\
In our setting, $M_{2} = M_{1}B + E$ can be interpreted as an approximate factor model with $M_{2} = F \Lambda + E$, where $F = U_{r}$ and $\Lambda = \Sigma V_{r}^T$, given that $M_{1}B = U_{r}\Sigma V_{r}^T$. However, despite this approximate factor structure, factor-based models are not well-suited for our problem. The decomposition used in Integrative Factor Regression will introduce a correlated nuisance parameter $F$, which may obscure the true effect of $M_{1}$. Furthermore, when $M_{1}(I,B)$ lacks spiked eigenvalues, selecting a appropriate number of factors becomes challenging in Factor-Adjusted Regularized Regression. This often results in choosing an excessively large number of factors, distorting the contribution of $M_{1}$ and leading to suboptimal model performance. Factor-based models impose structural assumptions that may not align well with the dependencies present in multi-modal data. The risks associated with obscuring meaningful relationships, introducing highly correlated nuisance parameters, and improperly selecting the number of factors make these methods less effective in our problem setting. A more detailed discussion of this issue is provided in the Supplementary Material Appendix \ref{appB}.

\subsection{Estimation}
To fit model (2), we first need to estimate the coefficient matrix \( \hat{B} \) that captures the relationship between \( M_{1} \) and \( M_{2} \). Several well-established reduced-rank regression methods can be utilized for this estimation. For instances, row-sparse reduced-rank regression (~\cite{Chen2012-bq}), sparse orthogonal factor regression (~\cite{8685192}), and multivariate reduced-rank linear regression (~\cite{chen2013reduced}) provide different sparsity assumptions for estimating \( \hat{B} \).  In general, the reduced-rank regression problem can be formulated as the following optimization problem:
\begin{align}
&  \min_{U,D,V} \| M_{2} - M_{1} UDV^{T} \|^{2}_{F} + \lambda_{1} \| D \|_{1} + \lambda_{2} \rho_{a}(UD) + \lambda_{3} \rho_{b}(VD) \notag \\
& \quad s.t. \quad U^{T} U = I,V^{T} V = I, B = UDV^{T}  
\end{align}
where the $U^{T} U = I,V^{T} V = I$ are introduced for identifiable purpose. $\rho_{a}$ and $\rho_{b}$ are penalty functions. They can be entry-wise $L_{1}$ norm or row-wise $L_{2,1}$ norm. $\lambda_{1}, \lambda_{2}, \lambda_{3}$ are the tuning parameters that control the magnitude of regularization. This framework generalizes several well-known methods: Row-sparse Reduced-Rank Regression when \( \lambda_{1} = \lambda_{3} = 0 \) and \( \rho_{a} = \| \cdot \|_{2,1} \); Multivariate Reduced-Rank Linear Regression when \( \lambda_{1} = \lambda_{2} = \lambda_{3} = 0 \); Sparse Orthogonal Factor Regression when all tuning parameters are nonzero. The tuning parameters and rank $r$ are chosen based on the GIC (\cite{fan2013tuning}). With the fitted model above, we could obtain the coefficient matrix $\hat{B}$ and residual term $\hat{E}$. \\
Next, we estimate the $P$ by the inner product of the first $r$ left singular vectors $U^{'}_{r}$ from $M_{1} \hat{B}$. Denote the estimation as $\hat{P}$. Then, transformed $M_{1}$ and $M_{2}$ can be estimated based on previous procedures. Once the transformed matrices are obtained, we estimate \( \hat{\beta_{1}} \) and \( \hat{\beta_{2}} \). This is done by solving the following penalized least squares problem:
\begin{align}
&  \min_{\beta_{1}, \beta_{2}, \gamma_{2}} \| Y - \hat{M^{*}_{1}} \beta_{1} - \hat{M^{*}_{2}}\beta_{2} - U_{r} \gamma_{2}\|^{2} + \lambda \rho(\beta_{1}) + \lambda \rho(\beta_{2})
\end{align}
where $\rho$ is a generic penalty function including the L1 norm, adaptive Lasso, MCP, and SCAD penalties. $\lambda$ is a tuning parameter that controls the regularization power on both $\beta_{1}$ and $\beta_{2}$. \\
For MVOPR with three modalities, the estimation of \( \beta_{1}, \beta_{2}, \) and \( \beta_{3} \) follows a similar penalized least squares approach:
\begin{align}
  \min_{\beta_{1}, \beta_{2}, \beta_{3}, \gamma_{1}, \gamma_{2}} & \| Y - \hat{M^{*}_{1}} \beta_{1} - \hat{M^{*}_{2}} \beta_{2} - \hat{M^{*}_{3}} \beta_{3} - U_{1}\gamma_{1} - U_{2}\gamma_{2}\|^{2}  \notag \\
& \quad + \lambda \rho(\beta_{1}) + \lambda \rho(\beta_{2})+ \lambda \rho(\beta_{3}) 
\end{align}
where $\hat{M^{*}_{1}} = (I - \hat{P_{1}})M_{1}$, $\hat{M^{*}_{2}} = (I - \hat{P_{1}})\hat{E_{2}}$, and $M^{*}_{3} = \hat{E_{3}}$. $U_{1}$ and $U_{2}$ are the left singular vectors with non-zero singular values of $M_{1} (\hat{B}_{2,1}\hat{B}_{3,1})$ and $E_{2} \hat{B}_{3,2}$.

\section{Numerical analysis}
\subsection{Variable selection on two modalities}
\subsubsection{$E$ with identity covariance matrix}
To assess the performance of MVOPR in comparison to other methods, we carry out some simulations under different noise levels on $\epsilon_{1}$ and $\epsilon_{2}$. In this simulations, suppose there are two modalities $M_{1}$ and $M_{2}$ with 300 features and 200 observations. $M_{1}$ is generated from multivariate normal distribution $MVN(0_{p},\Sigma_{M_{1}})$ with identity covariance matrix. Assume $M_{2}$ is connected with $M_{1}$ through a low-rank row sparse coefficient matrix $B$ with 95\% rows as zeros with rank $r=1$. Response $Y$ is associated with both $M_{1}$ and $M_{2}$ through $\beta_{1}$ and $\beta_{2}$. $\beta_{1}$ and $\beta_{2}$ are generated with 290 zeros and 10 non-zeros coefficients. The values of non-zero coefficients are sampled from uniform distribution $U(1,2)$. We fix the signal-to-noise ratio (SNR) to be 100 for $\epsilon_{1}$.By varying the SNRs of $\epsilon_{2}$, we compare the variable selection performance of each model by AUC. Each  AUC is calculated based on a 100 length grid of $\lambda$ which controlling strength of sparsity. We conduct each simulation experiment 100 times for one SNR of $\epsilon_{2}$. 

\begin{figure}[H]
\centering\includegraphics[width=13cm]{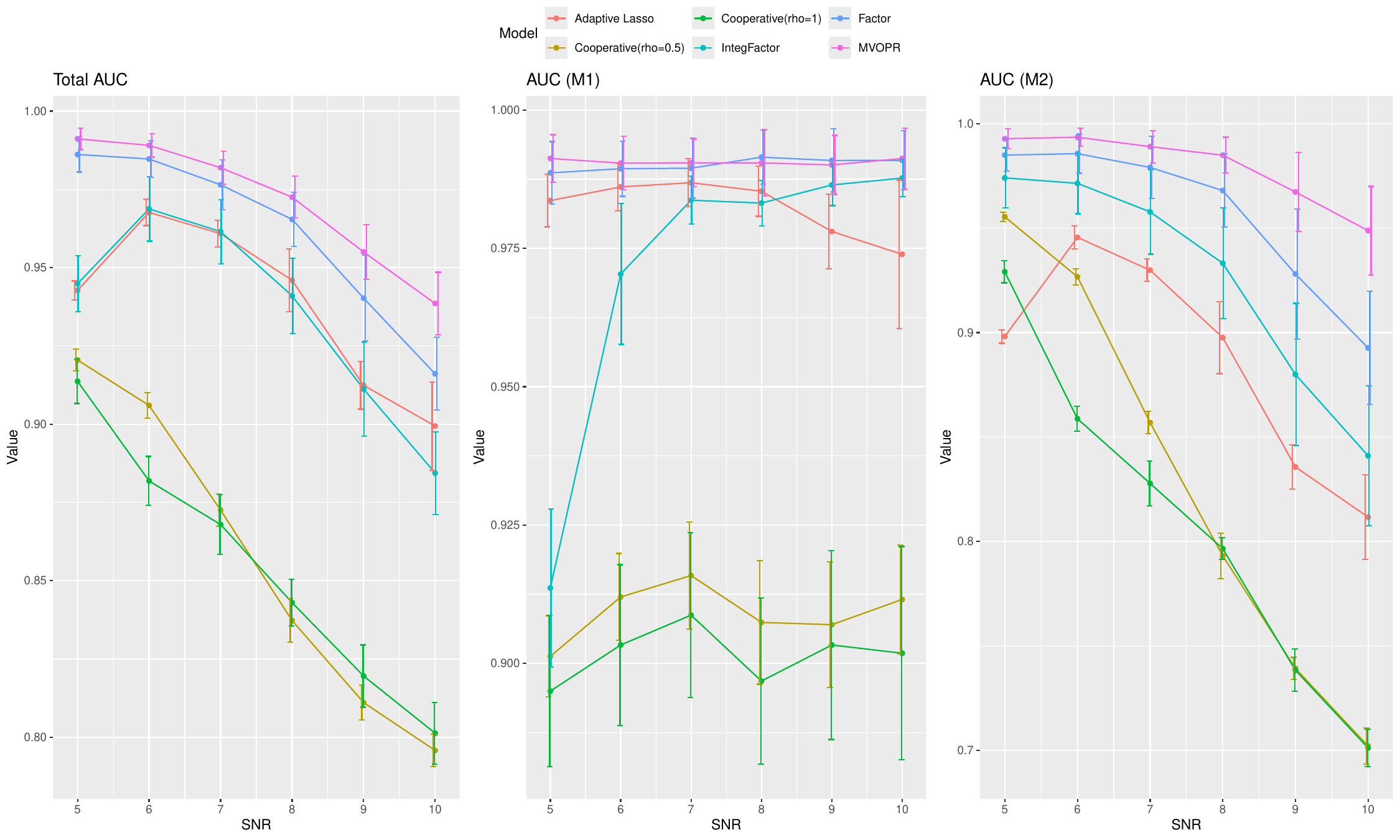}
\caption{The AUC for each model by varying the SNR of $\epsilon_{2}$ from 5 to 10}
\label{Figure.1}
\end{figure}
MVOPR outperforms other methods in terms of AUC across the entire range of SNR values (Figure.\ref{Figure.1}). Factor-Adjusted Regularized Regression exhibits comparable performance to MVOPR in scenarios with low SNR. However, as the SNR increases, which corresponds to stronger correlations between $M_{1}$ and $M_{2}$, MVOPR demonstrates clear superiority over Factor-Adjusted Regularized Regression. Especially, MVOPR has evident benefits on variable selection for $M_{2}$ when SNR is large. It shows the ability of MVOPR to better integrate information across modalities which allows it to maintain high AUC values even under more challenging conditions. In contrast, Factor-Adjusted Regularized Regression appears to struggle under high SNR conditions, likely due to its reliance on factor decomposition. Moreover, other competing methods, such as Integrative Factor Regression and Cooperative learning method, show declining performance as the SNR increases. These methods appear to be less effective in maintaining robust performance when faced with strong inter-modality correlations, highlighting the advantage of MVOPR in such scenarios.\\
Two alternative simulations are designed to show factor-based model may not be the ideal model to account for the inter-modality correlations. In the first simulation, $M_{1}$ and $M_{2}$ are generated from multivariate normal distribution with identity covariance matrix with 50 and 300 features. Each has 200 samples. Suppose $M_{2}$ is connected with $M_{1}$ through a low-rank row sparse coefficient matrix $B$ 70\% rows as zeros with rank $r=9$. $\beta_{1}$ and $\beta_{2}$ only has 10 non-zero coefficients separately. The second simulation is used to showcase the performance of MVOPR under low-dimensional data compared. $M_{1}$ and $M_{2}$ are generated from multivariate normal distribution with identity covariance matrix with 200 samples and 50 features. Low-rank row sparse coefficient matrix $B$ has 50\% rows as zeros with rank $r=3$. $\beta_{1}$ and $\beta_{2}$ has 25 non-zero coefficients separately. The SNR for $\epsilon_{1}$ in both simulations are fixed as 100. 
\begin{figure}[H]
\centering\includegraphics[width=13cm]{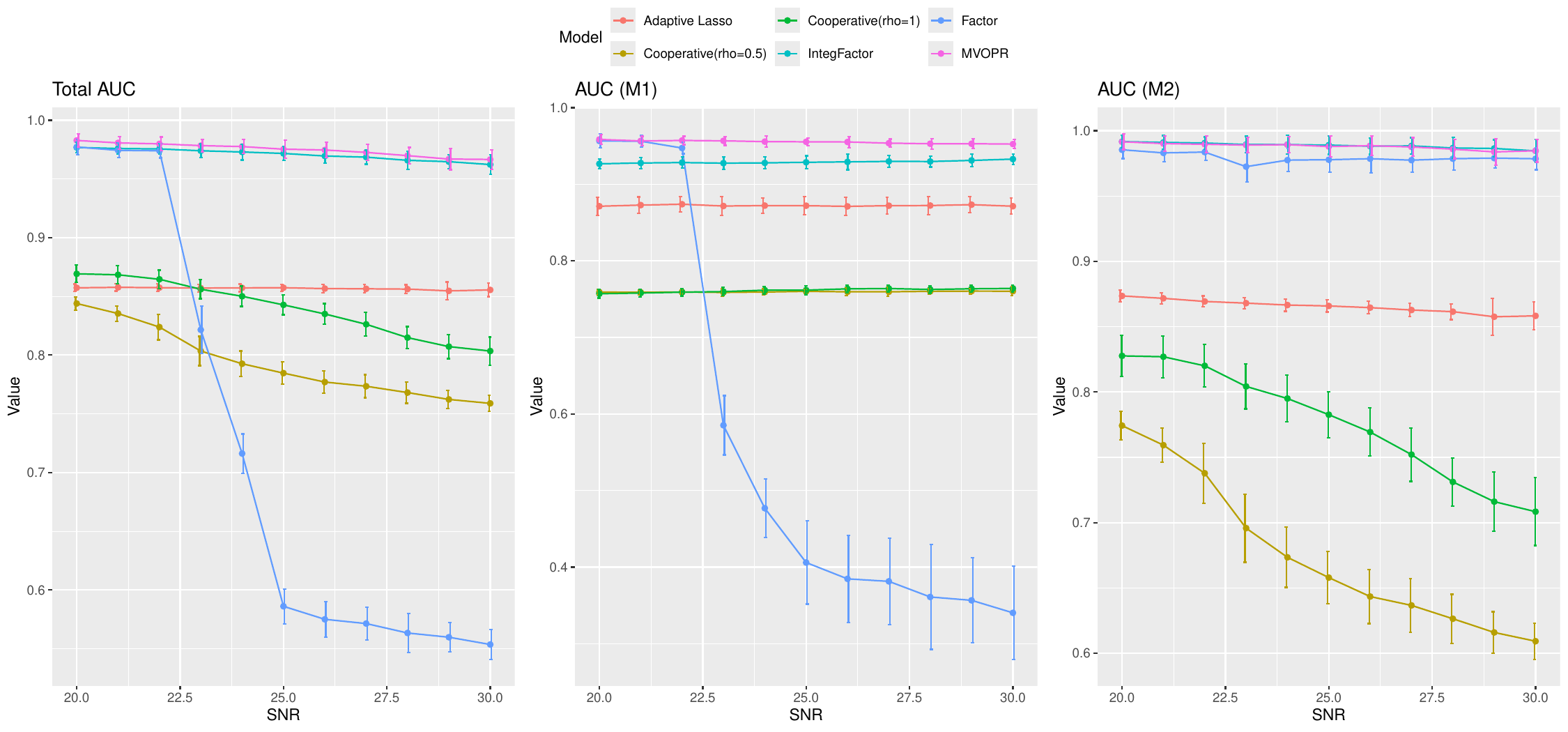}
\caption{The AUC for each model by varying the SNR of $\epsilon_{2}$ from 20 to 30 when $M_{1}$ and $M_{2}$ has different number of features. }
\label{Figure.2}
\end{figure}
\begin{figure}[H]
\centering\includegraphics[width=13cm]{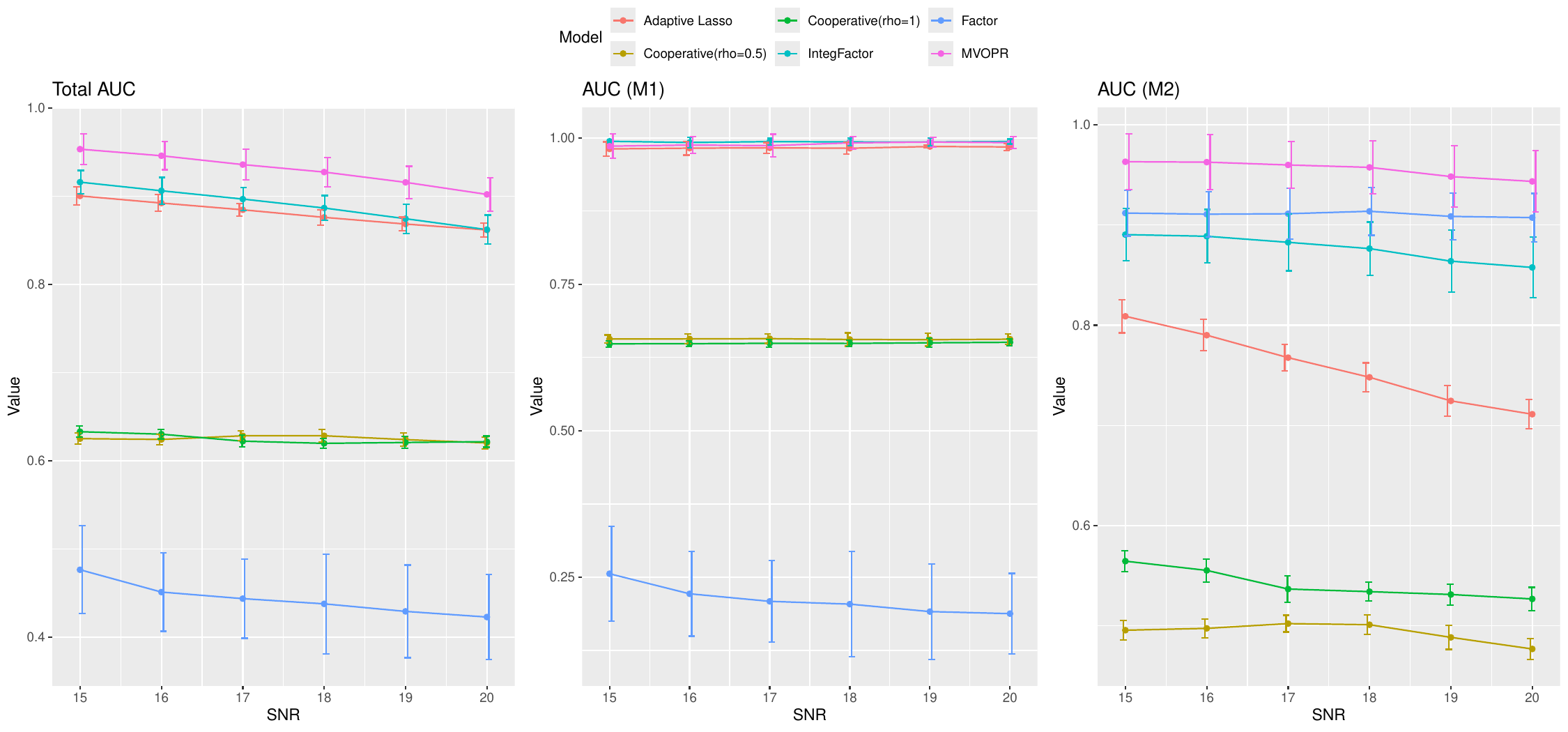}
\caption{The AUC for each model by varying the SNR of $\epsilon_{2}$ from 15 to 20 when $M_{1}$ and $M_{2}$ has 50 features respectively. }
\label{Figure.3}
\end{figure}
Under strong correlations between $M_{1}$ and $M_{2}$, MVOPR performs consistently well in both $M_{1}$ and $M_{2}$ variable selections compared to other method (Figure.\ref{Figure.2}, \ref{Figure.3}). Factor-Adjusted Regularized Regression shows comparable performance to MVOPR when SNR is smaller than 22. However, as inter-modality correlation become stronger, its variable selection ability for $M_{1}$ declines dramatically. This decline is likely attributed to the selection of excessively large number of factors, which overwhelms the meaningful signal and reduces its performance in isolating relevant variables from $M_{1}$. This problems become more obvious under low-dimensional simulation, where $(M_{1},M_{2})$ are more likely to be decomposed to a structure with excessive factors (Figure.\ref{Figure.3}). Integrative Factor Regression consistently underperforms in selecting variables for $M_{1}$, even when it selects zero number of factors for this modality. It can be attributed to its correlated nuisance variables with $M_{1}$, which disrupts the variable selection for predictors. This limitation underscores the difficulty of maintaining a balance between factor decomposition and effective variable selection in the presence of strong inter-modal correlations (Figure.\ref{Figure.2}).

\subsection{Misspecified Case}
\subsubsection{$E$ with correlated structure}
\begin{figure}[H]
\centering\includegraphics[width=14cm]{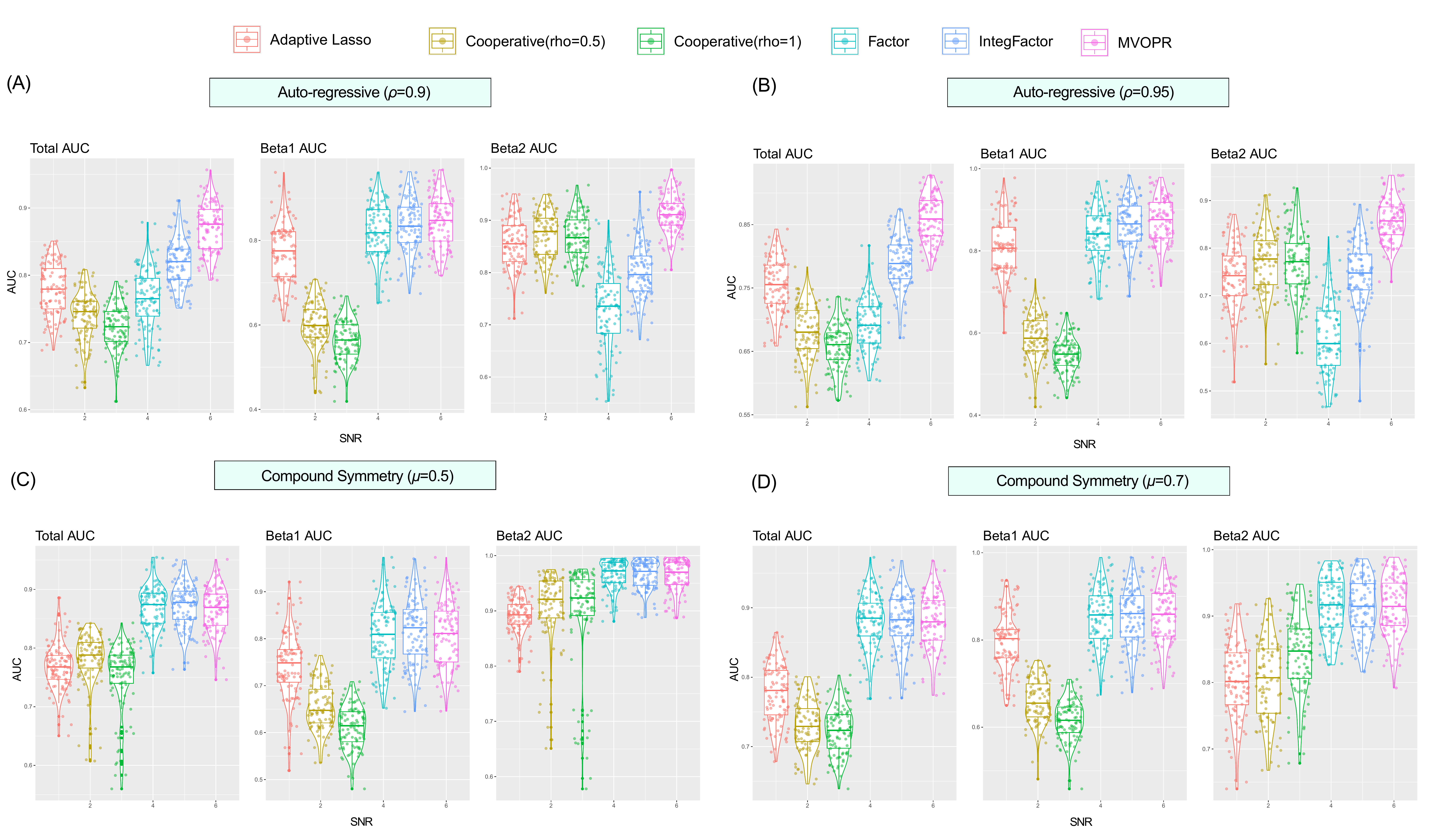}
\caption{The AUC of each model when $\epsilon_{1}$ has certain correlated structure. Fig.2.A-B showcase the AUC for each model under Auto-Regressive (AR1) covariance pattern. Fig.2.C-D showcase the AUC for each model under Compound Symmetry (CS) covariance pattern. }
\label{figure.2}
\end{figure}

In real world settings, $E$ may not always have independent covariance structure. To verify whether MVOPR can still works under this misspecified case, we consider two covariance patterns including auto-regressive (AR1) and compound symmetry (CS). In the simulations below, we generate two modalities $M_1$ and $M_2$ while each has 300 features and 200 observations. $M_{1}$ is generated from $MVN(0_{p},\Sigma_{M_{1}})$ with identity covariance matrix.$M_{2}$ is associated with $M_{1}$ through a low-rank row sparse coefficient matrix $B$ 50\% rows as zeros with rank $r=1$. Response $Y$ is associated with both $M_{1}$ and $M_{2}$ through $\beta_{1}$ and $\beta_{2}$. $\beta_{1}$ and $\beta_{2}$ are generated with 90 zeros and 10 non-zeros coefficients. The absolute values of non-zero coefficients are sampled from uniform distribution $U(1,2)$. The SNR for $\epsilon_{1}$ and $\epsilon_{2}$ are fixed to be 3 and 5. We compare the variable selection performance of each model by AUC.  In AR1 case, $E$ is generated from $MVN(0_{q},\Sigma_{\rho})$. The diagonal elements of $\Sigma_{\rho}$ are 1 with $cov(\epsilon^{i}_{2},\epsilon^{i}_{2}) = \rho^{|i-j|}$. $\rho=0.9$ and $\rho=0.95$ conditions are included. The results are shown in Figure.\ref{figure.2}.A and Figure.\ref{figure.2}.B. Under this misspecified case, MVOPR still achieves a higher AUC than other methods. In compound symmetry case, $E$ are generated from a $MVN(0_{q},\Sigma_{\mu})$. The diagonal elements of $\Sigma_{\mu}$ are 1 with $cov(\epsilon^{i}_{2},\epsilon^{i}_{2}) = \mu$. We test the performance of each model under $\mu=0.7$ and $\mu=0.9$ condition. The results are shown in Figure.\ref{figure.2}.C and Figure.\ref{figure.2}.D. Under this condition, both MVOPR and factor-based models performs well compared to adaptive lasso. 

\subsubsection{Null Experiment: No inter-modality correlation}
\begin{figure}[H]
\centering\includegraphics[width=13cm]{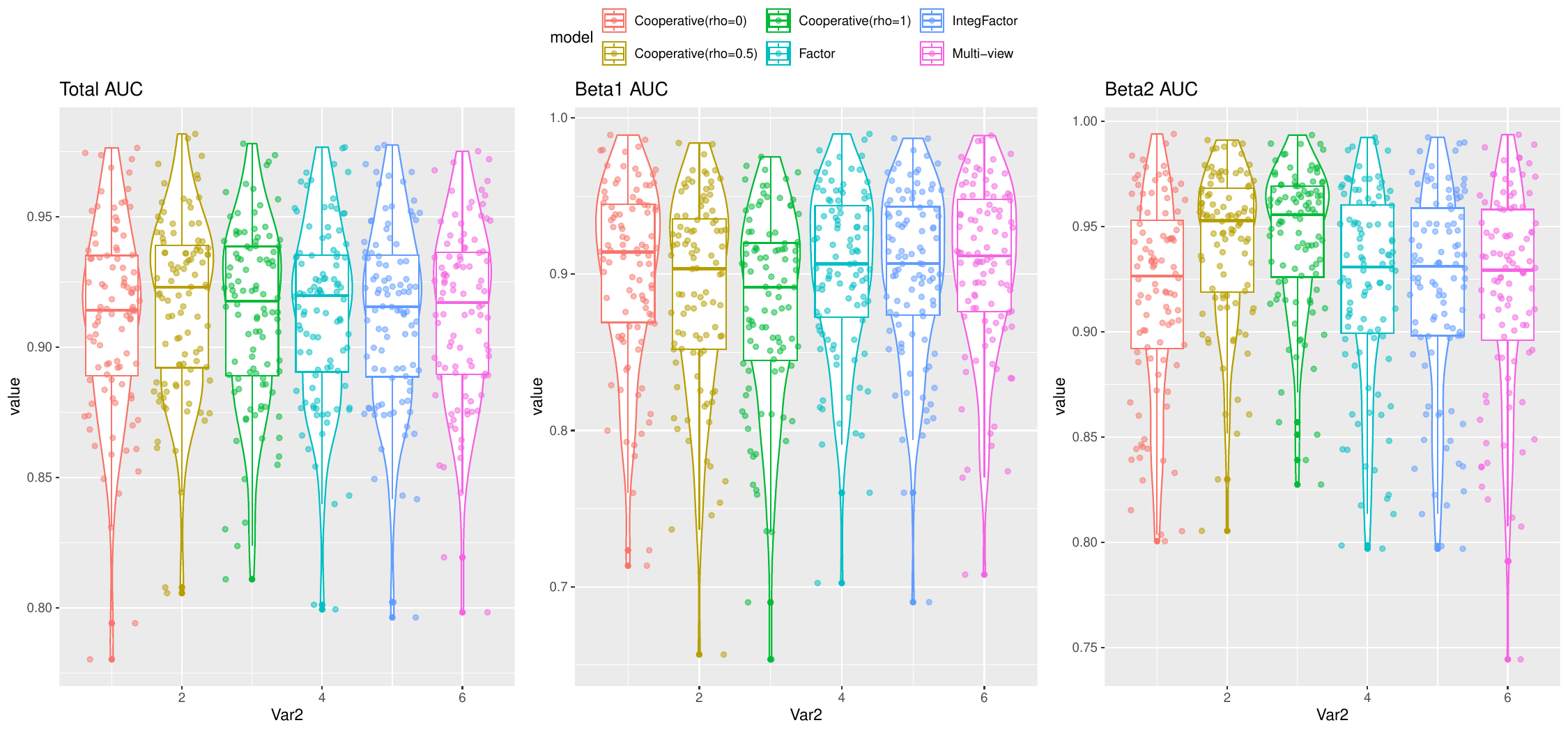}
\caption{The AUC of each model when both $M_{1}$ and $M_{2}$ have diagonal covariance matrix without the $M_{2} = M_{1}B$ assumption}
\label{figure.3}
\end{figure}

\begin{figure}[H]
\centering\includegraphics[width=13cm]{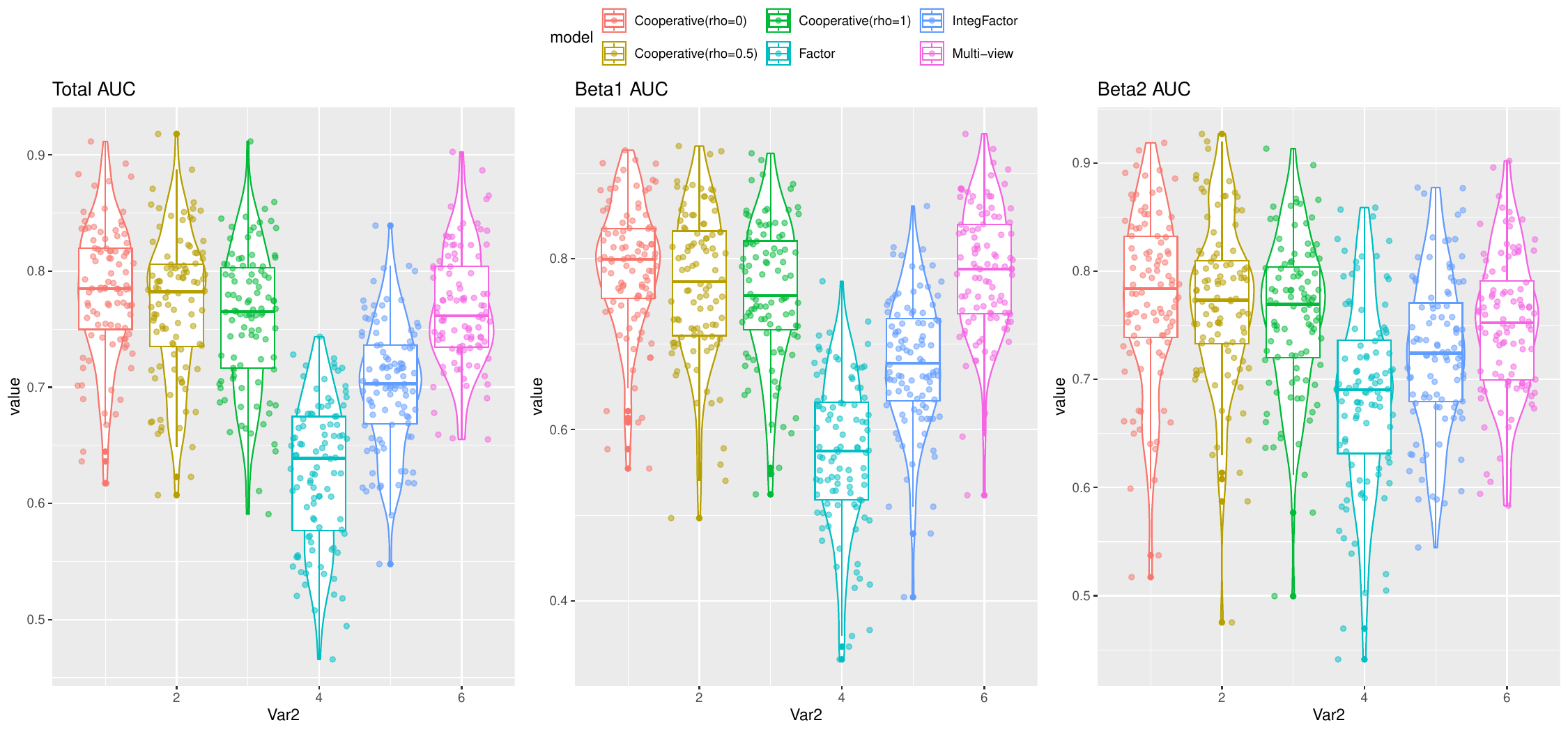}
\caption{The AUC of each model when both $M_{1}$ and $M_{2}$ have autoregressive ($\rho=0.9$) covariance matrix without the $M_{2} = M_{1}B$ assumption}
\label{figure.4}
\end{figure}

To evaluate whether MVOPR can still perform well when the $M_{2} = M_{1}B + \epsilon$ assumption is missing, we generate both $M_{1}$ and $M_{2}$ from $MVN(0_{p},\Sigma_{M_{1,2}})$ independently. In this simulation, we treat $\Sigma_{M_{1,2}}$ as identity or auto-regressive ($\rho=0.9$) covariance matrix. Suppose both $M_{1}$ and $M_{2}$ have 100 features and 200 samples. $Y$ is associated with $M_{1}$ and $M_{2}$ based on $\beta_{1}$ and $\beta_{2}$ which are generated based on the same rule in the previous section 3.2.1. \\
Based on the results in Figure.\ref{figure.3} and Figure.\ref{figure.4}, we notice that four models perform similarly under $\Sigma_{M_{1,2}} = I$. Meaning that even when the unidirectional assumption is missing, MVOPR can still work and share similar performance to other methods. However, if $\Sigma_{M_{1,2}}$ follows an auto-regressive ($\rho=0.9$) covariance pattern, factor-based models exhibit weaker performance compared to adaptive Lasso and MVOPR. This may be attributed to the covariance structure of each modality, as the absence of spiked eigenvalues hinders the effectiveness of factor decomposition.

\subsection{Simulation for Multi-modalities}
To evaluate the empirical performance of MVOPR on multi-modalities condition, we consider three modalities case from model (7). Suppose there are three modalities $M_{1}$, $M_{2}$, and $M_{3}$. Each modality has the same number of variables $p=p_{1} = p_{2} = p_{3} = 100$ with 100 observations. $B_{1}$, $B_{2}$, and $B_{3}$ are three low-rank coefficient matrix with rank $r_{1}=3, r_{2} = r_{3} = 1$. $B_{1}$, $B_{2}$, and $B_{3}$ are dense matrices with no row-wise sparsity. $Y$ is the response variable associated with $M_{1}$, $M_{2}$, and $M_{3}$.  \\
The estimations of $\hat{B_{1}}$, $\hat{B_{2}}$, and $\hat{B_{3}}$ are based on Multivariate Reduced-Rank Regression. $E_{2}$ and $E_{3}$ are generated from $MVN(0_{p}, \Sigma)$, while $\Sigma$ follows the identity. In this simulation, the SNRs for $E_{2}$ and $E_{3}$ are fixed to be 10 and 20. We also consider a misspecified case where $E_{2}$ and $E_{3}$ has correlated covariance structures. 

\begin{figure}[H]
\centering\includegraphics[width=12cm]{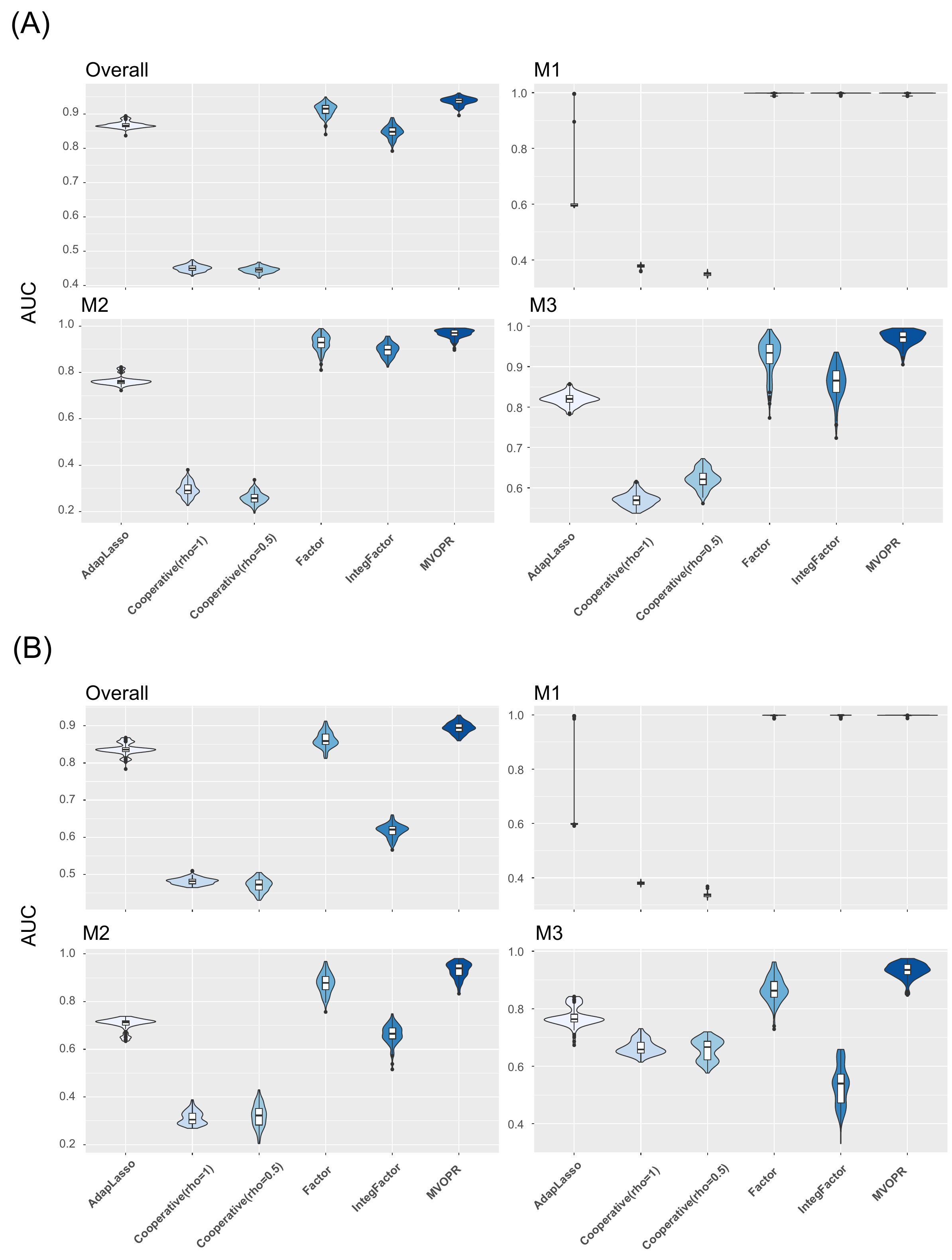}
\caption{AUC of the variable selection in $M_{1}$, $M_{2}$, and $M_{3}$. (A) The AUC distributions of MVOPR and other methods when $E_{2}$ and $E_{3}$ have identity covariance; (B) The AUC distributions of MVOPR and other methods when $E_{2}$ and $E_{3}$ have AR1 covariance}
\label{figure.5}
\end{figure}
Based on the Figure.\ref{figure.5}, we find that MVOPR outperforms other methods in terms of overall AUC, AUC in $M_{2}$, and AUC in $M_{3}$. MVOPR achieves the highest mean AUC values in these categories, indicating its superior performance in multi-modal integration. Among the competing methods, factor-based models show some improvement in overall and AUC for $M_{2}, M_{3}$ compared to Adaptive Lasso. However, their performance is worse than MVOPR. Specifically, Integrative Factor Regression exhibits a notable decline in performance when $E_{2}$ and $E_{3}$ share a correlated covariance structure. This result suggests that factor-based models may struggle to capture the intricate inter-modalality correlations. For Cooperative Learning, these models perform worse than adaptive Lasso, indicating that the agreement penalty may not always bring benefits to variable selection in these settings. The simulation results reveal the robustness and superiority of MVOPR in handling complex multi-omics settings. Even in misspecified scenarios, MVOPR consistently outperforms competing methods, demonstrating its reliability and effectiveness in capturing intricate correlations.

\section{Real data analysis}
\subsection{CAARS Data Analysis}
We conduct the MVOPR model on the CAARS data, collected from 55 patients. This dataset contains two omics layers: microbiome and metabolome. The study aims to understand how the omics datas influence the continuous eosinophil count. To reduce the dimensionality of microbiome and metabolome, we only select the metabolites which has top 200 variance. 139 microbiome are aggregated to 31 family levels. Then, we normalize the microbiome by centered log ratio transformation and the metabolome data is centered and scaled. We use the square root of continuous eosinophil count as response.\\
Multivariate reduced rank regression is applied to estimate the coefficient matrix $\hat{B}$ with the metabolome as a response and the microbiome as predictor. Original omics datasets are transformed based on the $\hat{B}$ and residuals $\hat{E}$. To analyze the association between the square root of continuous eosinophil count and transformed data, L1 peanlty is used for variable selection. Using a leave-one-out sample to qualify the robustness for variable selection. Out-sample MSE and stability indicators are used to qualify the performance of each model.  If one feature is selected with non-zero coefficient among 85\% iterations, it will be considered as a selected feature. Stability indicators are defined as follows: suppose the $i th$  set of variables selected by the model during the $i th$ iterations of leave-one-out sample as $S_{i}$. $S_{j}$ and $S_{i}$ are paired to calculate the stability. The two pairs are represented as $i$ and $j$ while $i \neq j$. Here are three stability indicators: Jaccard similarity coefficient, Otsuka–Ochiai coefficient, and Sørensen–Dice coefficient (\cite{kwon2023stability}). 

\begin{equation*}
\small
\begin{aligned}
Jaccard(S_{i}, S_{j}) &= \frac{|S_{i} \cap S_{j}|}{|S_{i} \cup S_{j}|} \quad & 
Ochiai(S_{i}, S_{j}) &= \frac{|S_{i} \cap S_{j}|}{\sqrt{|S_{i}| \cdot |S_{j}|}} \quad &
Dice(S_{i}, S_{j}) &= \frac{2|S_{i} \cap S_{j}|}{|S_{i}| + |S_{j}|} 
\end{aligned}
\end{equation*}
In the results (Table.\ref{table.1}), MVOPR achieves the lowest Mean Squared Error (MSE) of 177.73 and highest Jaccard similarity (0.66), Otsuka–Ochiai coefficient (0.75), and Sørensen–Dice coefficient (0.74). Those results reflect its robustness in variable selection and model stability. Additionally, MVOPR successfully selects five features. Notably, these selected features also appear as non-zero coefficients in traditional Lasso regression during some iterations. However, their selection frequencies are low, and their confidence intervals cross zero, indicating a lack of significance and stability in the traditional Lasso approach. In contrast, MVOPR not only identifies these features consistently, but also provides stronger evidence of their significance. The Factor-Adjusted Regularized Regression and Integrative Factor Regression models have relatively worse stability and MSE compared to MVOPR. These results suggest that although factor-based models may partially account for the within-modality correlations, they may not fully leverage the inter-modality correlation as effectively as MVOPR.

\begin{table}[H]
\begin{tabular}{lccccc}
\hline
\multicolumn{1}{c}{Models}             & MSE    & Jaccard  & Otsuka–Ochiai & Sørensen–Dice & Selected Features \\ \hline
Multi-view regression                  & 177.73  & 0.66   & 0.75  & 0.74        & 5                 \\
Lasso regression                       & 223.44  & 0.27    & 0.35  & 0.32      & 0                 \\
Factor Regression & 188.59  & 0.44   & 0.62  & 0.56        & 1                 \\
Integrative Factor Regression          & 414.49 & 0.41   & 0.53  & 0.47       & 1                 \\ \hline
\end{tabular}
\label{table.1}
\end{table}

\begin{figure}[H]
\centering\includegraphics[width=14.5 cm]{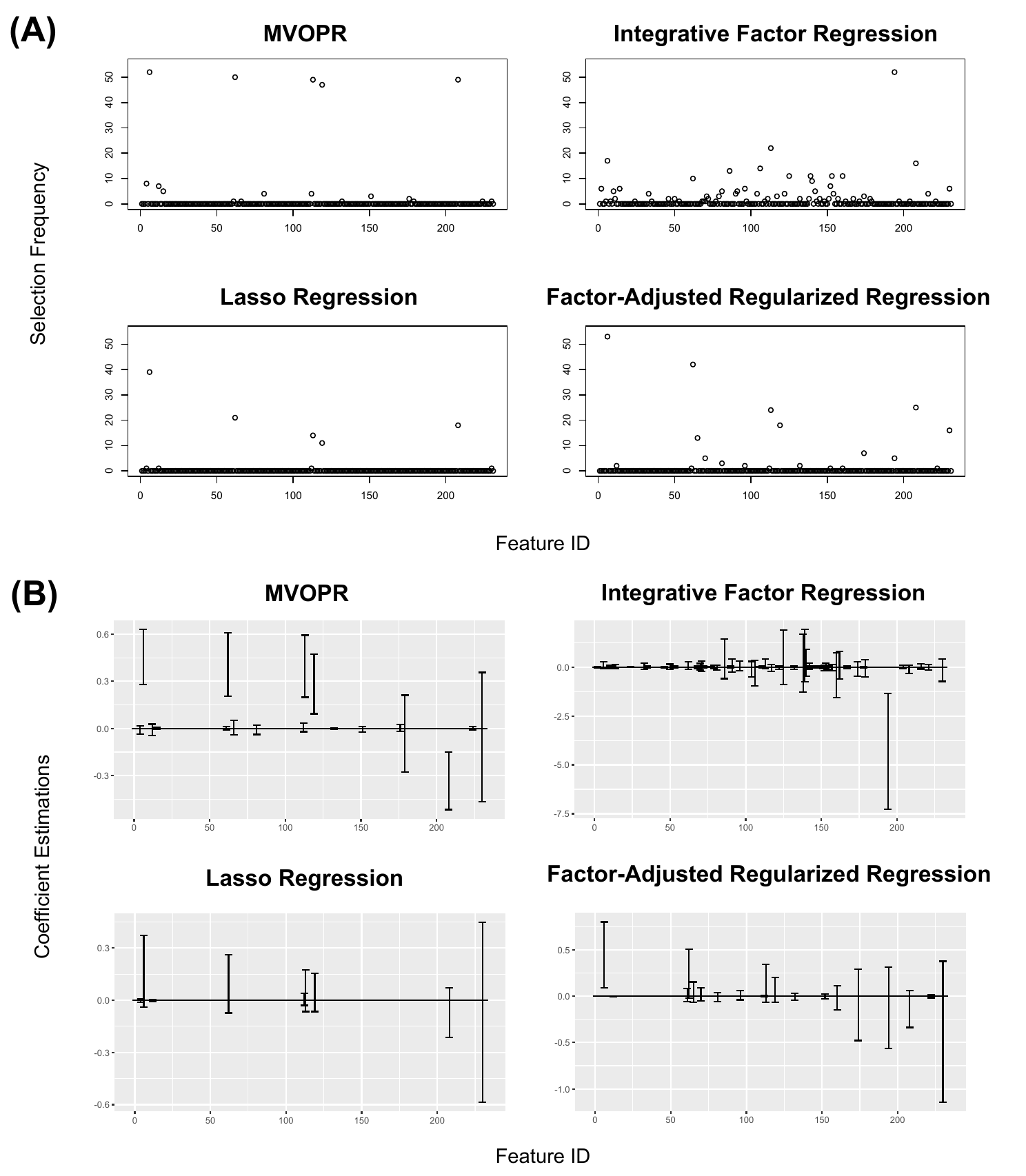}
\caption{MVOPR for CAARS Data Analysis. (A) Selection frequency Microbiome (ID: 1-31) and Metabolome (ID: 32 - 231); (B) Confident Intervals for Coefficients Estimations; (C) Pearson Correlation Matrix between Selected Microbiome and Metabolome}
\label{figure.6}
\end{figure}

\begin{figure}[H]
\centering\includegraphics[width=14.5 cm]{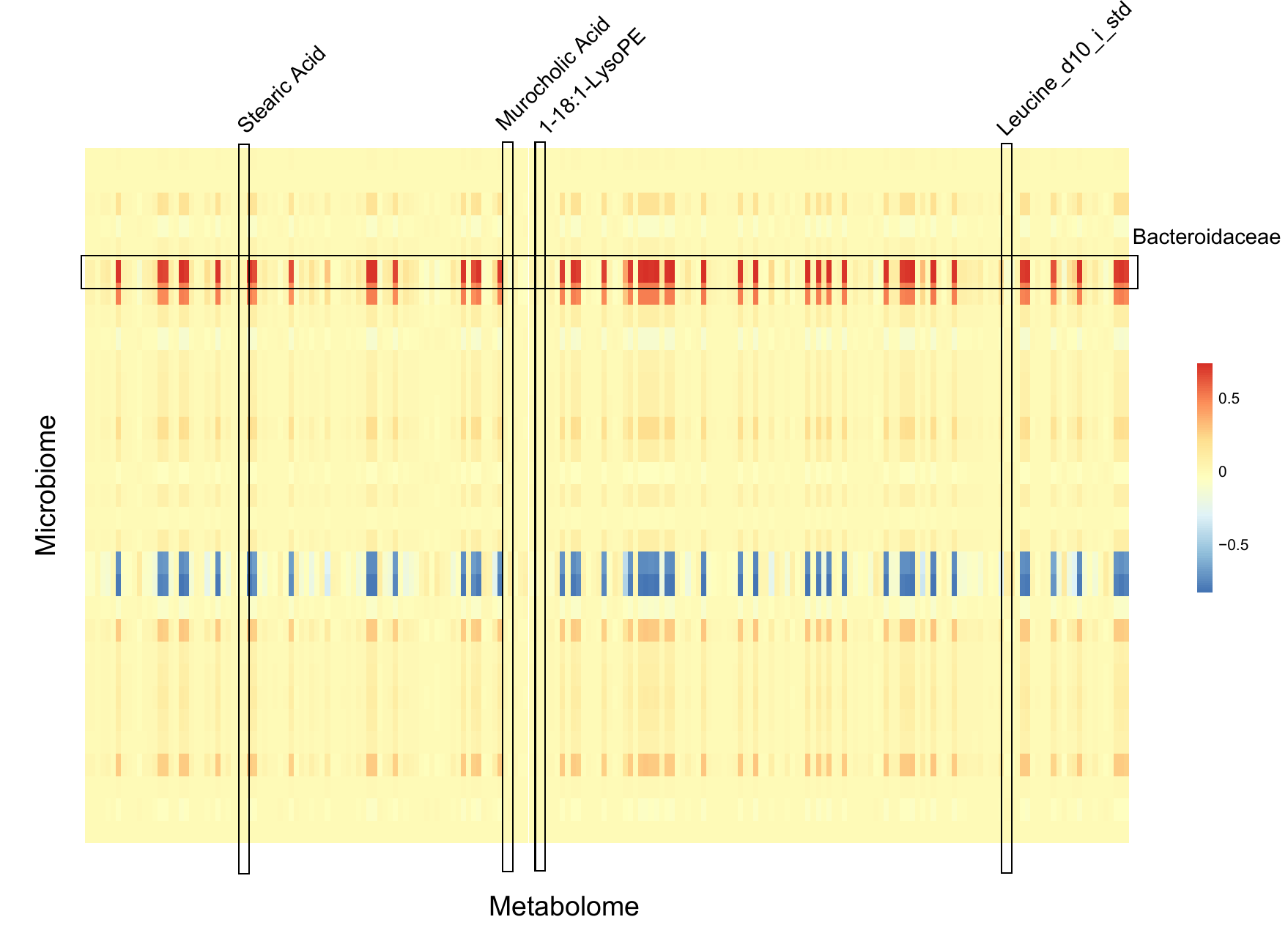}
\caption{Pearson Correlation Matrix between Selected Microbiome and Metabolome within MVOPR}
\label{figure.6}
\end{figure}

In MVOPR, \textit{Bacteroidaceae} family is consistently selected as a nonzero coefficient across 52 iterations, showing a positive average effect on the square root-transformed continuous eosinophil count. This result aligns with previous research suggesting that an increased relative abundance of \textit{Bacteroidaceae} is associated with asthma development (\cite{zimmermann2019association}). Within \textit{Bacteroidaceae}, the genera \textit{Bacteroides} plays a particularly important role in asthma pathophysiology. Some studies have identified \textit{Bacteroides} as a key microbial component in asthma progression (\cite{mahdavinia2023gut,aslam2024link,fiuza2024gut}). Among the four metabolites, stearic acid (tearic\_acid\_duplicate\_2), murocholic acid (murocholic\_acid\_duplicate\_2), 1-18:1-LysoPE (lyso\_pe\_18\_1\_9z\_0\_0\_duplicate\_2), and leucine (leucine\_d10\_i\_std). Stearic acid has been previously identified as a biomarker for asthma, showing elevated levels in asthma patients. Studies by ~\cite{tao2017metabonomics, tao2019urine} demonstrated that stearic acid exhibited excellent performance in distinguishing asthma patients from healthy controls. In our analysis, stearic acid also exhibited a positive correlation with the square root of the continuous eosinophil count, aligning with these findings. Muricholic acid has been linked to asthma in obesity models. A study by ~\cite{barosova2023metabolomics} found that obese mice with induced asthma had significantly higher muricholic acid levels compared to obese control mice. In our results, muricholic acid was positively correlated with eosinophilic inflammation, supporting its role in asthma pathophysiology. LysoPE (lysophosphatidylethanolamine) is a member of the lysophospholipids which is a large subclass of phospholipids. In previous studies under inflammatory diseases, researchers found that the signals of lysophospholipids was associated with the Chronic Obstructive Pulmonary Disease (~\cite{madapoosi2022lung}). In our analysis, leucine is identified as a significant metabolite associated with eosinophilic inflammation. Consistent with prior findings, higher leucine levels have been reported in asthmatic individuals with elevated exhaled nitric oxide (FeNO > 35), a biomarker indicative of eosinophil-driven inflammation (~\cite{comhair2015metabolomic}). This suggests that leucine may play a role in asthma pathophysiology, particularly in individuals with active eosinophilic reaction. The interplay between \textit{Bacteroides} and these metabolites is further supported by lipidomic analyses. Notably, differences in lipid profiles among \textit{Bacteroides} are largely driven by variations in plasmalogens, glycerophosphoinositols, and certain sphingolipids. These lipidomic distinctions may influence immune responses and inflammation, providing insight into the mechanisms by which \textit{Bacteroides} species contribute to asthma pathogenesis (~\cite{ryan2023membrane, ryan2023lipidomic}). Above all, MVOPR demonstrates strong performance in real data analysis, effectively identifying key microbial and metabolic features associated with eosinophilic inflammation in asthma. By selecting the \textit{Bacteroidaceae} family and relevant metabolites, MVOPR aligns well with established biological findings, highlighting its ability to capture meaningful microbiome-metabolome interactions. These results highlight the robustness and reliability of MVOPR in modeling complex multi-omics relationships, making it a powerful tool to uncover biomarkers in asthma.

\section{Discussion}
The MVOPR model presents a novel approach for multi-omics data integration by using the orthogonal projection framework to handle the correlated predictors, enhancing variable selection. Our model is effective under the unidirectional assumption, aligning well with the inherent biological pathways such as the Central Dogma of Molecular Biology. Traditional methods, such as Lasso-based regression and factor-based models, struggle in multi-omics settings due to the strong within- and inter-modality correlations. MVOPR effectively addresses these challenges by leveraging the unidirectional assumptions between omics layers and employing an orthogonal projection framework to mitigate multicollinearity problems. \\
Based on the results from simulations and real data analysis, MVOPR showcases superior performance over other competing methods. Unlike factor-based models, which require an approximate factor structure on predictors, MVOPR successfully eliminates redundant dependencies while preserving meaningful signals for variable selection. Even in scenarios where the inter-modality correlation assumption is violated, MVOPR maintains competitive performance, outperforming other methods. This suggests that MVOPR generalizes well beyond ideal conditions, making it a reliable tool for real-world applications. \\
However, in cases where the model is severely misspecified, such as incorrectly assuming directionality, performance of MVOPR can be affected. For instance, if the true causal direction is from $M_{1}$ to $M_{2}$, but a model with reverse direction is fitted ($M_{2}$ to $M_{1}$), the estimated coefficient matrix $\hat{B}$ may not be well-constructed, leading to poor projections and inaccurate variable selection. To ensure proper unidirectional modeling, a strong understanding of the latent relationships between modalities is crucial. This can be established through biological knowledge, such as the Central Dogma of Molecular Biology, or causal inference that helps to determine the correct directionality before model fitting. \\ 
In current analysis, MVOPR operates within a linear regression framework. However, some biological systems are inherently nonlinear and hierarchical, often involving complex interactions between different omics layers. Future extensions of MVOPR could incorporate nonlinear model, such as kernel-based methods or deep-learning approaches, to capture these intricate dependencies more effectively. \\
When applying MVOPR to the CAARS dataset, we successfully identify microbial and metabolic markers linked to eosinophilic inflammation in asthma. Notably, MVOPR select some biomarkers which aligns with prior research. Compared to competing approaches, MVOPR demonstrate higher model stability and lower mean squared error (MSE) in real-data analysis. Traditional methods such as Lasso regression and factor-based models failed to maintain consistent variable selection across iterations. In contrast, MVOPR achieved higher stability indicators (Jaccard, Otsuka–Ochiai, and Sørensen–Dice coefficients), suggesting improved stability in biomarker identification. \\
MVOPR represents a advancement in multi-omics variable selection, providing a robust, interpretable, and biologically relevant framework for multi-view data integration. By successfully mitigating within- and inter-modality correlations, MVOPR allows for more precise biomarker discovery, particularly in complex diseases such as asthma. As multi-omics datasets continue to develop, MVOPR offers a powerful and stable method for integrative analysis, providing novel framework for personalized medicine and targeted therapeutic strategies.

\begin{acks}[Acknowledgments]
The authors would like to thank the anonymous referees, an Associate
Editor and the Editor for their constructive comments that improved the
quality of this paper.
\end{acks}

\begin{funding}
The first author was supported by NSF Grant DMS-??-??????.

The second author was supported in part by NIH Grant ???????????.
\end{funding}

\begin{appendix}

\section{Extension to multiple modalities}\label{appA}
\subsection{Three modalities case}
For three modalities case, we could first transform $M_{2}$ and $M_{3}$ into their residuals forms $E_{2}$ and $E_{3}$. The model (8) would be rewriten as below: 
\begin{equation*}
Y = M_{1} \beta_{1} + E_{2} \beta_{2} + E_{3} \beta_{3} + M_{1} (B_{2,1} \beta_{2} + B_{3,1} \beta_{3}) + M_{2} B_{3,2}\beta_{3} + \epsilon_{1}
\end{equation*}
while $M_{2}$ could be further decomposed to $M_{1}B_{2,1} + E_{2}$. Therefore, 
\begin{equation*}
Y = M_{1} \beta_{1} + E_{2} \beta_{2} + E_{3} \beta_{3} + M_{1} (B_{2,1} \beta_{2} + B_{3,1} \beta_{3} + B_{2,1}B_{3,2} \beta_{3}) + E_{2} B_{3,2}\beta_{3} + \epsilon_{1}
\end{equation*}
Since two nuisance variable $M_{1} (B_{2,1}, B_{3,1})$ and $E_{2} B_{3,2}$ are correlated with predictors $M_{1}$ and $E_{2}$, we need to project those predictors to the orthogonal subspace. Suppose $M_{1} (B_{2,1}, B_{3,1}) = U_{1} \Sigma_{1} V_{1}^{T}$ and $E_{2} B_{3,2} = U_{2} \Sigma_{2} V_{2}^{T}$ with rank $r_{1}$ and $r_{2}$. Two projection matrices are $P_{1} = U_{1}U_{1}^{T}$ and $P_{2} = U_{2}U_{2}^{T}$. Based on the projection, we have: 
\begin{equation*}
Y = (I - P_{1})M_{1} \beta_{1} + (I - P_{2})E_{2} \beta_{2} + E_{3} \beta_{3} + U_{1}\gamma_{1} +U_{2}\gamma_{2} + \epsilon_{1} 
\end{equation*}
In this form, we will have mutually uncorrelated predictors and nuisance variables in the regression. Below, we make a brief discussion about the assumptions on independence between predictors and nuisance Variables: 

1. $E_{3} \perp\!\!\!\perp (I - P_{1})M_{1}$ and $E_{3} \perp\!\!\!\perp (I - P_{2})E_{2}$ hold since $E_{3} \perp\!\!\!\perp M_{1}$ and $E_{3} \perp\!\!\!\perp E_{2}$. \

2. $(I - P_{2})\epsilon_{2} \perp\!\!\!\perp (I - P_{1})M_{1}$ holds since $E_{2} \perp\!\!\!\perp M_{1}$. \

3. $(I - P_{1})M_{1} \perp\!\!\!\perp U_{1,r}$ and $(I - P_{2})E_{2} \perp\!\!\!\perp U_{2,r^{'}}$ hold since the projection matrix $P_{1},P_{2}$ are orthogonal to their complements $(I - P_{1}),(I - P_{2})$.\

4. $(I - P_{2})E_{2} \perp\!\!\!\perp U_{1,r}$ and $E_{3} \perp\!\!\!\perp U_{1,r}$ since $E_{2} \perp\!\!\!\perp M_{1}$ and $E_{3} \perp\!\!\!\perp M_{1}$. \

5. $(I - P_{1})M_{1} \perp\!\!\!\perp U_{2,r^{'}}$ and $E_{3} \perp\!\!\!\perp U_{2,r^{'}}$ since $M_{1} \perp\!\!\!\perp E_{2}$ and $E_{3} \perp\!\!\!\perp E_{2}$.

\subsection{Multiple modalities case}
For multi-omics data with $k$ modalities, we need to determine the order for each modality. Based on Central dogma of molecular biology, genomics will generally serve as the first modality which has the ability to influence all the downstream elements. Proteomics or metabolomics may serve as the last modality which can be regularized by upstream elements. Any omics between the first and last modality will serve as intermediate modality such as transcriptome. After we have the sequential information for multi-omics, we could transform each modality except to their residual forms first. 
\begin{align*}
Y & = M_{1} \beta_{1} + \sum_{j=2}^{k} E_{j} \beta_{j} + M_{1}B_{1}^{*}\gamma_{1} + \sum_{i=2}^{k-1}E_{i} B_{i}^{*}\gamma_{i} + \epsilon_{1}
\end{align*}
where $B_{1}^{*} = (B_{2,1},B_{3,1},...,B_{k,1})$. For any $2 \leq i \leq k-1$, $B_{i}^{*} = (B_{i+1,i},B_{i+2,i},...,B_{k,i})$. To remove the correlation between predictors and nuisance variables, we next project each predictors to the orthogonal subspace. Suppose we have SVD for each nuisance variable: 
\begin{align*}
M_{1}B_{1}^{*} = U_{1}\Sigma_{1}V_{1}^{T} \quad E_{i} B_{i}^{*} = U_{i}\Sigma_{i}V_{i}^{T}
\end{align*}
Assume $U_{1},U_{2},...,U_{k-1}$ has rank $r_{1}, r_{2}, ..., r_{k-1}$. Projection matrix $P_{1} = U_{1}U_{1}^{T}, P_{2} = U_{2}U_{2}^{T},..., P_{k-1} = U_{k-1}U_{k-1}^{T}$. Then, the final model for MVOPR will be: 
\begin{align*}
Y = P_{1}^{\perp}M_{1} \beta_{1} + \sum_{i=2}^{k-1} P_{i}^{\perp}E_{i} \beta_{i} + E_{k}\beta_{k} + \sum_{j=1}^{k-1} U_{j}\gamma_{j}^{*} + \epsilon_{1}
\end{align*}
The transformed modalities are mutually uncorrelated to each other in the regression. 

\section*{Connection to factor based model}\label{appB}
Assume $M_{1} \in \mathbb{R}^{n \times p}$ and $M_{2}\in \mathbb{R}^{n \times q}$ are two omics data. Let $Y$ denotes the response variable. Suppose $Y$ is associated with $M_{1}$ and $M_{2}$ by $\beta_{1} \in \mathbb{R}^{p}$ and $\beta_{2} \in \mathbb{R}^{q}$. The interplay between $M_{1}$ and $M_{2}$ can be captured by a low-rank coefficient matrix $B$ with rank $r$. $E$ and $\epsilon_{1}$ are the error matrix and vectors. 
\begin{equation*}
Y = M_{1} \beta_{1} + M_{2} \beta_{2} + \epsilon_{1}
\end{equation*}
\begin{equation*}
M_{2} = M_{1} B + E
\end{equation*}
Suppose $B$ matrix has a SVD as $B = U_{B} \Sigma_{B} V_{B}^T$. Therefore, $M_{2}$ could be expressed based on an approximate factor model. $F = M_{1} U_{B}$ and $\Lambda = \Sigma_{B} V_{B}^T$. This structure aligns with the scenarios for Integrative Factor Regression and Factor-Adjusted Regularized Regression.  
\begin{align*}
M_{2} & = M_{1} U_{B} \Sigma_{B} V_{B}^T + E \\
& = F \Lambda + E 
\end{align*}
Suppose $M_{1}$ doesn't follow approximate factor model structure. In Integrative Factor Regression, the factor decomposition for $(M_{1},M_{2})$ will become $(M_{1},E)$ with factors $F$ as nuisance parameter. However, since the factors are given by $F = M_{1} U_{B}$, it implies that $F$ is a linear combination of $M_{1}$ and is therefore highly correlated with it. When we fit the regression $Y = M_{1} \beta_{1} + E \beta_{2} + F \gamma + \epsilon_{1}$, the true contribution of $M_{1}$ will be obscured by the correlated nuisance parameter $F$. When $M_{1}$ has some spiked eigenvalues and could be approximated by factor models, similar problem will still exist. Suppose $M_{1} = F_{1} \Lambda_{1} + U_{1}$, the decomposition of $(M_{1},M_{2})$ will become $(U_{1},E)$ with nuisance parameters $(F_{1},F)$. Since $F = M_{1} U_{B} = F_{1}\Lambda_{1}U_{B} + U_{1}U_{B}$, $F$ will still be correlated to $F_{1}$ and $U_{1}$. \\
In Factor-Adjusted Regularized Regression, the matrix $M = (M_{1},M_{2})$ is treated as a whole and decomposed accordingly. Since M follows the decomposition $M = M_{1}(I,B)+(0,E)$ which does not perfectly align with the model’s assumption, the selection of the number of factors will be affected. When $B$ has a rank $r$ that is closed or equal to $p$, $M$ could be decompose to $F\Lambda+(0,E)$ with $p$ factors. In this case, the transformed $M_{1}$ will be nearly zero, as most of its information is absorbed by the factors. Similar issue will happen when $M_{1}(I,B)$ lacks spiked eigenvalues, leading to difficulties in distinguishing factor structure. This increases the risk of selecting an excessively large number of factors, potentially distorting the factor adjustment process.

\end{appendix}



\bibliographystyle{imsart-nameyear} 
\bibliography{sample.bib}

\end{document}